\documentclass[twocolumn,prd,floatfix,nofootinbib]{revtex4}
\usepackage{amsmath}
\usepackage{graphicx}
\usepackage{dcolumn}
\usepackage{bm}
\usepackage{amssymb}
\usepackage{latexsym}

\def\be{\begin{equation}}
\def\ee{\end{equation}}
\def\ba{\begin{eqnarray}}
\def\ea{\end{eqnarray}}

\begin{document}

\title{Rotating a Curvaton Brane in a Warped Throat}

\author{Jun Zhang$^{a}$\footnote{Email: zhangjun408@mails.gucas.ac.cn}}
\author{Yi-Fu Cai$^{b}$\footnote{Email: caiyf@ihep.ac.cn}}
\author{Yun-Song Piao$^{a}$\footnote{Email: yspiao@gucas.ac.cn}}

\affiliation{${}^a$ College of Physical Sciences, Graduate
University of Chinese Academy of Sciences, Beijing 100049, China}

\affiliation{${}^b$ Institute of High Energy Physics, Chinese
Academy of Sciences, P.O.Box 918-4, Beijing 100049, P.R.China}

\begin{abstract}
In this paper we study a curvaton model obtained by considering a probe
anti-D3-brane with angular motion at the bottom of a KS throat
with approximate isometries. We calculate the spectrum of
curvature perturbations and the non-Gaussianities of this model.
Specifically, we consider the limit of relativistic rotation of
the curvaton brane which leads to a small sound speed, and thus it
can be viewed as an implementation of the DBI-curvaton mechanism. We find
that the primordial power spectrum is nearly scale-invariant while
the non-Gaussianity of local type is sizable and that of
equilateral type is usually large and negative. Moreover, we study
both the theoretical and observational constraints on this model,
and find that there exists a sizable allowed region for the phase
space of this model.
\end{abstract}

\maketitle

\section{Introduction}

The curvaton mechanism \cite{Lyth:2001nq} is an interesting
proposal for explaining the observed scale-invariant primordial
density perturbations in the framework of inflation (and some
pioneering works on this mechanism have been considered in Refs.
\cite{Mollerach:1989hu, Linde:1996gt, Enqvist:2001zp}). In this
scenario, at least two fields are required, in which the curvaton
field is subdominant during inflation but becomes dominant
after the inflaton decays. When the curvaton is subdominant it
provides entropy (iso-curvature) perturbations during inflation,
and afterwards, these entropy perturbations can be converted to
curvature perturbations when the curvaton starts to dominate the
universe. After the curvaton decays, the universe enters the
standard thermal history, and then the primordial curvature
perturbations lead to the formation of the large scale structure
of our universe\cite{Lyth:2002my, Lyth:2003ip}.

Recently, a curvaton scenario realized in the frame of stringy
inflationary models was presented in Ref. \cite{Li:2008fma}. As
advocated in Ref. \cite{Dvali:1998pa}, a system of D- and
anti-D-branes in a warped throat can be made use of to provide an
inflationary stage in the early universe\cite{Kachru:2003si}, and
its dynamics is described by a Dirac-Born-Infeld (DBI) action
in the relativistic limit\cite{Silverstein:2003hf,
Chen:2004gc}. Therefore, the curvaton mechanism realized in this
background is the so-called ``DBI-curvaton", where the curvaton
field can be interpreted as a moduli parameter of a D-brane moving
in a warped throat\cite{Huang:2007hh}. However, it is widely
recognized that a brane inflation model of DBI-type often suffers
from problems of backreaction both from the relativistic brane
moving in the warped throat\cite{Silverstein:2003hf, Chen:2005ad,
Easson:2007dh} and from the inflationary background (see
\cite{Chen:2008hz} for an extended discussion), and so is often
unable to provide a enough long inflationary stage required by
cosmological observations\cite{Bean:2007hc, Peiris:2007gz,
Bean:2007eh}. The original DBI-curvaton model also suffers from
the same problem as that of DBI inflation.

To circumvent this problem, one has to finely tune the precise
shape of the potential of the inflaton in the presence of
supersymmetrically embedded D7-branes and an anti D3- brane
localized at the tip of the warped conifold cone, and therefore
the resulting model is rather delicate as studied in Refs.
\cite{Baumann:2007np, Krause:2007jk, Baumann:2007ah}. Another
approach to easing this tension suggests that the motion of a
probe brane along the angular coordinates in the internal space
can be taken into account to prolong the inflationary
period\cite{DeWolfe:2007hd} which relaxed the theoretical
constraints from the gravitational backreaction, and so extends
the phase space of brane inflation\cite{Easson:2007dh}.

In this paper, we study a new DBI-type curvaton model established
on the scenario of multiple brane inflation proposed in Refs.
\cite{Cai:2008if, Cai:2009hw}. We suggest that a probe
anti-D3-brane with angular momentum on the tip of an approximately
isometric warped throat can play the role of a curvaton field.
Specifically, we consider the well-known Klebanov-Strassler (KS)
throat \cite{Klebanov:2000hb} which is a nonsingular deformed
conifold with its moduli being stabilized by
fluxes\cite{Giddings:2001yu, Kachru:2003aw}. In this model, the
anti-D3-brane always sits at the bottom of the deformed throat due
to the sum of gravitational and gauge forces. Correspondingly its
radial position is nearly fixed and the angular motion can be
safely preserved during inflation.

Since the specific realization of this curvaton mechanism involves
at least two branes with the anti-D3-brane as the curvaton brane,
this model can be viewed as an application of the multi-brane
inflation model\cite{Cai:2008if, Cai:2009hw, Piao:2002vf}, and it
naturally possesses some features of the multi-brane inflation
model. For instance, the spectral index of iso-curvature
perturbations generated by the curvaton brane during inflation has
a slight red tilt deviating from scale-invariance, and is mainly
dependent on the slow-roll parameter characterizing the variation
of the Hubble parameter during inflation. At cubic order we find
the nonlinear perturbations of a curvaton brane model have
the combined features of a usual curvaton model and a
DBI-inflation one. The non-Gaussianity of local type nicely agrees
with the traditional result obtained in a general curvaton model;
however, the non-Gaussianity of equilateral type is amplified by
the small sound speed of the curvaton brane which is consistent
with the usual scenario of brane inflation.

Our paper is organized as follows. In Section II we briefly review
the dynamics of D- and anti-D-branes in the KS throat. In Section
III, we present the realization of the DBI-curvaton in the KS
throat and show that the curvaton brane can rotate in the
relativistic limit. In Section IV we calculate the curvature
perturbations and the non-Guassianities generated in this model.
The consistency relationships are analyzed in Section V. Section
VI contains discussion and conclusions.

\section{Brane dynamics in a warped throat}

To begin with, we consider a flux compactification of type IIB
string theory on an orientifold of a Calabi-Yau
three-fold\cite{Giddings:2001yu}. The geometry of this warped
deformed conifold is viewed as a solution to type IIB supergravity
with the presence of negative tension sources.
It can be explicitly constructed by taking a stack of D3-branes
with $M$ units of the Ramond-Ramond (RR) fluxes $F_3$ on the $S_3$
(A-cycle) and $K$ units of the Neveu-Schwarz-Neveu-Schwarz (NS-NS)
fluxes $H_3$ on the dual cycle (B-cycle) which is the $S_2$ times
a circle extended along the radial direction. This gives,
\begin{eqnarray}
 \frac{1}{2\pi \alpha'} \int_A F_3 = 2\pi M~, ~~~~
 \frac{1}{2\pi \alpha'} \int_B H_3 = -2\pi K~,  \label{MK}
\end{eqnarray}
where $\alpha'$ is related to the string mass scale by $\alpha' =
1/m_s^2$. When the fluxes are turned on, they warp the geometry of
the conifold and the background metric of the 10D spacetime can be
expressed as follows,
\begin{eqnarray}
 ds^2 = h^2 g_{\mu\nu} dx^{\mu} dx^{\nu} + h^{-2}
 (dl^2 + l^2 \tilde g_{mn} d{y}^m d{y}^n)~,
\end{eqnarray}
where the warping factor $h$ can be a function of the radius $l$
and internal coordinates $y^m$ generically. Additionally,
$g_{\mu\nu}$ is the metric of 4-dimensional spacetime and $\tilde
g_{mn}$ characterizes the internal space of the compact manifold.

In the above system, we have only introduced the D3-branes. The
geometry away from the tip of the conifold is an approximately
Anti-de-Sitter (AdS) form with
\begin{eqnarray}
h \simeq \frac{l}{L}~,
\end{eqnarray}
and the characteristic length scale $L$ is fixed by the flux
number as
\begin{eqnarray}
L^4 \simeq \frac{27}{4}\pi g_s MK \alpha'^2~,
\end{eqnarray}
where $g_s$ is the string coupling which has to satisfy $g_sM \gg
1$ to ensure the validity of the supergravity
description\cite{Aharony:1999ti}. In order to obtain a de-Sitter
(dS) solution which may be used to realize inflation, we can
introduce anti-D3-branes at the tip of the throat, which can lift
the AdS vacuum such as the KKLT mechanism\cite{Kachru:2003aw}.
Note that, the warping factor at the infrared (IR) end of the
throat takes
\begin{eqnarray}
 h_{IR} \simeq \exp \left(-\frac{2\pi K}{3 g_s M}\right)~, \label{h0}
\end{eqnarray}
and the IR cutoff of the radial coordinate $l_{IR} \simeq (g_s M
\alpha')^{1/2} h_{IR}$.

Now, we are interested in the dynamics of the branes in the KS
throat. It has been shown that the $\overline{D3}$-brane will sink
to the tip of the throat, while the D3-brane is almost free except
attracted by the $\overline{D3}$-brane only. To have a quick look,
we can write down the effective action for a D3-brane or an
$\overline{D3}$-brane \cite{Myers:1999ps},
\begin{equation}
 S = -T_3 \int d^4x \sqrt{-det[G_{\mu \nu}]} \pm T_3 \int d^4x (C_4)_{0123}~,
 \label{DBI-CS}
\end{equation}
with $G_{\mu \nu}$ being defined as
\begin{equation}
G_{\mu \nu} = G_{AB} \frac{\partial X^A}{\partial x^{\mu}}
\frac{\partial X^B}{\partial x^{\nu}}~,
\end{equation}
where $+$ is for D3-brane, $-$ for $\overline{D3}$, and the $T_3$
is the tension which is given by $T_3 =
\frac{m_s^4}{(2\pi)^3}g_s$. The equation of motion of the RR field
yields,
\begin{equation}
(C_4)_{0123} = \sqrt{-g} h^4~.
\end{equation}
So, if the transverse fluctuations of the brane vanish, we will
find that, for D3, these two contributions to the action cancel
exactly; however, they could sink $\overline{D3}$ to the bottom of
the throat.

As usual, if the KS throat is isometric, the warp factor $h$ is
independent of the angular coordinates ${y}^m$. However, in a
general case, there are some corrections which make the warp
factor dependent on angular coordinates. For example, since the
compact C-Y manifold cannot have exact continuous isometries, the
isometries of the bulk must be broken when the finite throat is
glued on this bulk, and then the warp factor becomes
angular-dependent. Since these effects are suppressed by powers of
the warp factor, the correction to the warp factor can be
described in terms of $\Delta(h^4) \sim h_{IR}^\beta f ({y})$ as
analyzed in Refs. \cite{Ceresole:1999zs, DeWolfe:2004qx,
Aharony:2005ez}. However, we shall keep in mind that,
$\Delta(h^4)$ can not be too large, otherwise the geometry of the
KS throat would be distorted seriously.

A nonperturbative effect which stabilizes the K$\ddot{a}$hler
moduli could bring a potential for the branes \cite{Berg:2004ek,
Baumann:2006th}. The form of this potential depends on the precise
embedding of the wrapped branes, and for embeddings which do not
admit supersymmetric vacua on the tip, this potential could be
angular-dependent. In Ref. \cite{DeWolfe:2007hd}, the authors have
provided an estimate of the nonperturbative effect which is
dominated by the warp factor in the case of the Kuperstein
embedding\cite{Kuperstein:2004hy} of the D7-brane, where the
angular degrees of freedom obtain a mass term,
\begin{equation}\label{massnp}
 m_{np}^2 \simeq \frac{h_{IR}^2}{g_s M \alpha'}
 \frac{\epsilon}{\mu}~,
\end{equation}
in which $\mu$ measures the minimal radial location reached by the
D7-brane, and $\epsilon \sim \mu h_{IR}^{3/2}$ characterizes the
deformation of the coinfold.

To combine the corrections mentioned above, we can obtain the
effective action of the $\overline{D3}$-brane as follows,
\begin{equation}\label{Action-g}
S = -T_3 \int d^4x \sqrt{-det[G_{\mu \nu}]} - T_3 \int d^4x
\sqrt{-g} \left( h^4 + V_{np} \right)~,
\end{equation}
where $V_{np}$ stands for the angular-dependent potential given by
the nonperturbative effects, and so is a function of the angular
coordinates $y^m$. Note that we have neglected the interactions
among the curvaton and other fields for simplicity. In addition,
even if the isometries of the throat are broken, the dynamics of
the brane along the radial direction are almost the same as that
in the isometric throat, unless the bulk effects can distort the
throat geometry drastically.

At the end of this section we would like to comment on the tadpole
condition which is a limit on the net brane charge. Since the
branes carry charges, they cannot be inserted into a compact space
optionally. The extra brane charges must be compensated by the
background, and thus the following condition has to be satisfied
\begin{equation}
\frac{\chi}{24} = N_{D3}-N_{\overline{D3}} +
\frac{1}{\kappa_{10}^2 T_3} \int_{C-Y} H_3 \wedge F_3~,
\end{equation}
where $\chi$ is the Euler number of the C-Y manifold, and
$N_{D3}$($N_{\overline{D3}}$) stands for the number of the
D3-brane ($\overline{D3}$-brane). In the usual case, $\chi$ is of
order $O(10^5)$ \cite{Klemm:1996ts}, and so it puts an upper bound
on the background charge number as $ MK \lesssim 10^5.
\label{euler}$

\section{The model of curvaton brane}

\begin{figure}[htbp]
\includegraphics[scale=0.5]{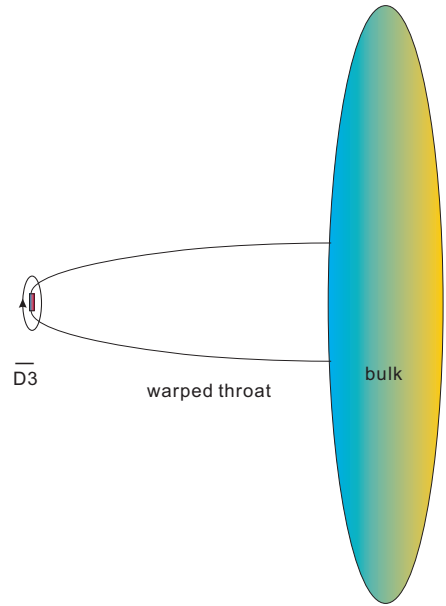}
\caption{ A rotating $\overline{D3}$-brane in a warped throat. In
a deformed KS throat, the branes are no longer pointlike in the
extra dimensions. The circle stands for the angular motion of the
brane at the infrared tip of the warped throat, $l_{IR}$. }
\label{fig:curvatonbrane}
\end{figure}

Within the background introduced in the previous section, now we
are able to construct the model of a DBI-curvaton from a
$\overline{D3}$-brane rotating in the KS throat. We consider a
$\overline{D3}$-brane with angular momentum at the tip of the KS
throat which is approximately isometric. As shown in the previous
section, this brane is localized in the radial direction. Note
that in Ref. \cite{Easson:2007dh} it was found that the angular
momenta of the branes could be inflated away if these branes were
used to drive inflation. In our construction, we assume that the
inflationary background is still driven by a D3-D7 system as
introduced in Refs. \cite{Baumann:2007np, Baumann:2007ah} (see
\cite{Herdeiro:2001zb, Dasgupta:2002ew} for earlier studies), the
radial coordinate of the $\overline{D3}$-brane is nearly fixed,
and there exists an external potential from the nonperturbative
effect of moduli stabilization which drives the curvaton brane to
rotate. Consequently, provided the external potential persists
sufficiently long, the angular momentum of the probe brane can
safely survive during inflation.\footnote{This is different from
Ref. \cite{Easson:2007dh} in which their driving force is still
almost radial and so leads to damping angular momenta. The
non-vanishing angular degree of freedom in a warped throat was
earlier applied to realize a bouncing
universe\cite{Germani:2006pf, Germani:2007uc}.} In this case, the
radial coordinate of the probe brane is around $l_{IR}$, and thus
we can study its dynamics along the angular directions at the tip
of the KS throat and its application in the curvaton mechanism.
The global scenario of our model is sketched in Fig.
\ref{fig:curvatonbrane}.

Substituting the induced metric on the $\overline{D3}$-brane into
its effective action (\ref{Action-g}), we can get a simplified
form,
\begin{eqnarray}\label{S2}
 S &\simeq& T_3 \int d^4x \sqrt{-g}
 [ - h^4 \sqrt{1+\frac{1}{h^4}
 ( l_{IR}^2 \tilde{g}_{mn} \partial_\mu{y}^m\partial^\mu{y}^n) } \nonumber\\
 &&- ( h^4 + V_{np} ) ]~,
\end{eqnarray}
where we have neglected the higher derivative terms on spatial
directions which are secondary on affecting the background
evolution and the primordial perturbations at large scales. Note
that, although the terms including higher spatial derivatives are
negligible at linear and even cubic order, they could affect the
nonlinear evolution of primordial perturbations at fourth order
and so could leave a significant imprint on the trispectra of this
model. This will be studied in following works in the near future.

Moreover, there are three angular directions of the $S_3$ tip, and
everyone is able to play the role of a curvaton field. We would
like to pick up the flattest direction, and define it as the ${y}$
coordinate, for simplicity. Consequently, we can define the
curvaton field as,
\begin{equation}
\sigma \equiv l_{IR} \sqrt{T_3} ~{y}~.
\end{equation}
Finally, the effective action of the curvaton brane can be
described by,
\begin{eqnarray}
S &\simeq& T_3 \int d^4x \sqrt{-g} \bigg\{ - h^4(\sigma)
 \sqrt{1+\frac{1}{h^4T_3} ( \partial_\mu\sigma\partial^\mu\sigma ) } \nonumber\\
  && - [ h^4(\sigma) + V_{np}(\sigma) ] \bigg\}~.
\end{eqnarray}
As usual, we define a sound speed parameter of the curvaton field
\begin{equation}\label{soundspeed}
c_s \equiv \sqrt{1-\frac{2}{h^4T_3}X}~,
\end{equation}
where $X$ is define as
$X\equiv-\frac{1}{2}(\partial_\mu\sigma\partial^\mu\sigma)$.
Further, we would like to define an effective potential of the
curvaton field as follows,
\begin{eqnarray}\label{potential}
 V(\sigma) \equiv T_3 \left(2 h^4 + V_{np} \right)~,
\end{eqnarray}
and thus the corresponding effective Lagrangian can be written as,
\begin{eqnarray}
 P \equiv T_3 h^4 \left( 1 - c_s \right) - V(\sigma)~,
\end{eqnarray}
of which the form is similar to that of the usual DBI model.

We notice that a similar scenario involving a slow-rotating
$\overline{D3}$-brane was considered in Ref.
\cite{Kobayashi:2009cm} and its action was expanded to quadratic
order under the assumption of the non-relativistic limit $c_s \sim
1$. In the current work, however, we are going to consider the
relativistic dynamics of the model of the curvaton brane with the
limit of $c_s\sim 0$. In a flat Friedmann-Robertson-Walker (FRW)
universe with the metric characterized by,
\begin{eqnarray}
ds_{FRW}^2=-dt^2+a^2(t)\delta_{ij}dx^idx^j~,
\end{eqnarray}
the equation of motion for the curvaton field $\sigma$ is given by
\begin{equation}\label{EOMB1}
\ddot\sigma+3H\dot\sigma-\frac{\dot{c_s}}{c_s}\dot\sigma-c_s
P_{,\sigma}=0
\end{equation}
where the dot represents the cosmic time derivative, and
$P_{,\sigma}$ denotes the derivative of $P$ with respect to the
scalar field $\sigma$. One may notice that, since the energy
density of the curvaton is secondary during inflation, the hubble
parameter $H$ is determined by the background D3-D7 inflationary
model.

Before studying the cosmological perturbations of this model, we
would like to study the evolution of its background equation under
the relativistic limit. In the case of $c_s \sim 0$, Eq.
(\ref{soundspeed}) yields
\begin{eqnarray}
\dot{\sigma}\simeq -h^2 \sqrt{T_3}~,
\end{eqnarray}
and it implies that the variation of the curvaton is roughly
$\Delta\sigma\sim -h^2 \sqrt{T_3}\Delta t$ for a fixed time
interval $\Delta t$. Besides, as mentioned in Section 2, provided
the geometry of the KS throat is not distorted too much, we still
have the infrared warp factor $h_{IR}$. Following Refs.
\cite{Cai:2008if, Cai:2009hw}, we have an ansatz with the
following form,
\begin{eqnarray}\label{ansatz}
 \sigma = - h_{IR}^2\sqrt{T_3}~t
 \left(1+\frac{\alpha}{(-t)^p}+\cdot\cdot\cdot\right)~,
\end{eqnarray}
where we have set $t\rightarrow-\infty$ at the beginning of
inflation. Moreover, Eq. (\ref{EOMB1}) can be rewritten as
\begin{equation}\label{EOMB2}
 \frac{d}{dt}(\frac{\dot{\sigma}}{c_s})
 +3H\frac{\dot{\sigma}}{c_s}
 -4T_3h_{,\sigma}h^3(1-c_s)
 +\frac{2h_{,\sigma}}{h}\frac{\dot{\sigma}}{c_s}
 +V_{,\sigma}=0~.
\end{equation}

To apply the ansatz (\ref{ansatz}) into Eq. (\ref{EOMB2}), we then
have the leading terms, which are the second term
\begin{eqnarray}
\frac{-3Hh_{IR}^2\sqrt{T_3}}{\sqrt{2\alpha(p-1)}(-t)^{-p/2}}~,
\end{eqnarray}
and the potential term
\begin{eqnarray}
m_{np}^2h_{IR}^2\sqrt{T_3}t~,
\end{eqnarray}
respectively. The nonperturbative mass term is given by Eq.
(\ref{massnp}), as introduced in Section 2. The others are
suppressed by the slow-varying factor $h_{,\sigma}$ and
$\frac{1}{Ht}$ (where $|Ht| \gg 1$ is required by inflation). By
matching the leading terms, we get the results $p=2$ and $\alpha =
\frac{9H^2}{2m_{np}^4}$. So we obtain the solution
\begin{eqnarray}\label{solution}
\sigma =
-h_{IR}^2\sqrt{T_3}~t\left(1+\frac{9H^2}{2m_{np}^4t^2}+\cdot\cdot\cdot\right)~,
\end{eqnarray}
for a relativistic curvaton brane rotating in a warped throat.

Notice that the value range of the curvaton is constrained by the
radius of the 3-circle at the tip. In the usual case, it restrict
the variation of the coordinate along the angular direction
$\Delta y$ to be less than $O(1)$. Consequently, to multiply the
hubble parameter on the solution (\ref{solution}), the angular
restriction brings a constraint on the efolding number of
inflation as follows,
\begin{eqnarray}
 {\cal N}_{\rm inf} \equiv
 \int Hdt \lesssim H\frac{(g_sM\alpha')^{\frac{1}{2}}}{h_{IR}}~.
\end{eqnarray}
In addition, the solution (\ref{solution}) gives a rolling-down
behavior for the curvaton field, which in the stringy frame shows
that the rotation angle of the curvaton brane becomes more and
more parallel to the 3-cycle at the tip, and the
angular-dependence will finally vanish. However, the quantum
fluctuation of the curvaton during inflation may be in conflict
with its classical variation so that hold the curvaton on the
plateau of its potential longer. Therefore, with the help of
quantum fluctuation there would be more angular modes preserved,
but the model may suffer from the potential tension of large
backreaction.

\section{Primordial perturbations of the curvaton brane}

In the above section we have studied the background evolution of
the curvaton brane during inflation. If our model indeed makes
sense to the physics of the early universe, it is necessary to
study the primordial perturbations generated by the curvaton
field. This provides a potential discriminant between numerous
curvaton models in the literature.

\subsection{Quantum fluctuations during inflation}

To begin with, we split the curvaton field $\sigma$ into
$\sigma(\textbf{x},t)=\sigma_0(t)+\delta\sigma(\textbf{x},t)$,
where $\sigma_0$ stands for the background field which is
homogeneous and isotropic, and $\delta\sigma$ stands for the
linear fluctuation which is caused by quantum effects.

Following the formulae of Ref. \cite{Cai:2009hw}, we study the
dynamics of quantum fluctuations in the spatially flat gauge,
and assume no couplings to the background inflaton. In this case,
we have the equation of motion describing the canonical
perturbation variable in Fourier space as follows,
\begin{eqnarray}\label{eomvk}
v_k''+(c_s^2k^2-\frac{z''}{z})v_k \simeq 0~,
\end{eqnarray}
where the prime denotes the derivative with respect to the
conformal time $\tau \equiv \int\frac{dt}{a}$. The canonical
perturbation variable is defined as
\begin{eqnarray}
v_k \equiv a c_s^{-3/2}\delta\sigma_k~,
\end{eqnarray}
and a background dependent function is introduced,
\begin{eqnarray}
z \equiv {a}{c_s^{-3/2}H^{-1}}\dot\sigma~,
\end{eqnarray}
in the above perturbation equation.

One may notice that Eq (\ref{eomvk}) has an asymptotic solution
when we neglect the last term $\frac{z''}{z}$, which implies
$|c_sk\tau|\gg1$, and it oscillates strongly like a sine function.
This feature coincides with the slow roll condition defined above,
which corresponds to the case in which the effective physical
wavelength is deep inside the so-call sound horizon
$\frac{c_s}{H}$. Therefore, the modes can be regarded as states in
the adiabatic Minkowski vacuum during the sub-Hubble regime.
Correspondingly, we can impose a suitable initial condition
\begin{eqnarray}\label{inicond}
v_k^{ini}=\frac{e^{-ic_sk\tau}}{\sqrt{2c_sk}}~,
\end{eqnarray}
which corresponds to the Bunch-Davies vacuum.

In the inflationary background, we have the following useful
approximate relation,
\begin{eqnarray}
\frac{z''}{z}\simeq\frac{2}{\tau^2}~,
\end{eqnarray}
which implies that the power spectrum of the field fluctuations
takes the form
\begin{eqnarray}\label{delta sigma}
\delta\sigma=\sqrt{P_{\delta\sigma}}\simeq\frac{H_*}{2\pi}~,
\end{eqnarray}
after they exit the sound horizon. The subscript `$*$' stands for
the sound horizon crossing time for the field fluctuation.

Since during inflation the Hubble parameter is nearly constant and
the field fluctuations are almost conserved after sound horizon
crossing, we are able to calculate the spectral tilt of the
primordial perturbations at the horizon crossing moment. This is
given by
\begin{eqnarray}\label{n sigma}
n_{\sigma}-1\equiv\frac{d\ln P_{\sigma}}{d\ln
k}=-2\epsilon_*+3s_*~,
\end{eqnarray}
where we have defined a series of slow-variation parameters,
\begin{eqnarray}
 \epsilon &\equiv& -\frac{\dot H}{H^2}~,\\
 s &\equiv& \frac{\dot c_s}{Hc_s}~.
\end{eqnarray}

Note that we have ignored the square and the higher derivatives of
the slow roll parameters in the above expressions. The
perturbation equation is derived under the assumption that the
coupling between the adiabatic and isocurvature modes is
negligible. A more detailed analysis of a similar scenario where
an inflation model is based on multiple branes has been given in
Ref. \cite{Cai:2009hw}. Thus we can naturally understand the
difference in the results obtained in the current model and those
obtained in \cite{Li:2008fma} where multiple moduli degrees of
freedom are involved in a single brane. Besides, to compare the
r.h.s. of Eq. (\ref{n sigma}) with that in the usual curvaton model,
we can see that the correction to the spectral index from the mass
term disappears in our model, because it is strongly suppressed by
a term proportional to $c_s^3$.

\subsection{Curvature perturbations}

To proceed, we calculate the curvature perturbations generated by
the curvaton field after inflation has ceased and the inflaton
field has already decayed to radiation. During this period, the
energy density of the universe is composed of $\rho_r$ which
denotes the radiation energy density and $\rho_{\sigma}$ which is
contributed by the curvaton. In this stage, the curvature
perturbations are generated from the isocurvature modes since the
pressure perturbations are non-adiabatic. This process ends when
the perturbations become adiabatic again, which corresponds to the
epoch of curvaton domination, or that of curvaton decay. The final
curvature perturbations can be calculated at the moment $H=\Gamma$
based on the assumption of perturbatively instantaneous reheating
of the curvaton.
In that case we can simply consider
the component curvature perturbations $\zeta_{\sigma}$ and
$\zeta_{\gamma}$ on slices of uniform curvaton density and
radiation density separately. According to the assumptions of the
curvaton mechanism, we can neglect the perturbations from
radiation and thus have the curvature perturbation expressed as,
\begin{eqnarray}
 \zeta=\frac{3(1+w_{\sigma}\rho_{\sigma})}{4\rho_{\gamma}+3(1+w_{\sigma})\rho_{\sigma}}
 \zeta_{\sigma}~,
\end{eqnarray}
where $w_{\sigma}\equiv p_{\sigma}/\rho_{\sigma}$ is defined as
the equation-of-state of the curvaton, and $\zeta_{\sigma}$ is
given by
\begin{eqnarray}
 \zeta_{\sigma}=\frac{\delta\rho_{\sigma}}{3(1+w_{\sigma})\rho_{\sigma}}~.
\end{eqnarray}

As emphasized in Section II, the nonperturbative effect which
stabilizes the K$\ddot{a}$hler moduli in the case of the
Kuperstein embedding yields a quadratic potential for the curvaton
brane as given in Eq. (\ref{potential}). Therefore, after
inflation the curvaton field arrives at its vacuum state and
starts to oscillate when $H=m_{np}$. At that moment, the sound
speed $c_s$ goes to $1$, and the energy density of the curvaton
field redshifts proportional to $a^{-3}$. So we have an
approximate equation-of-state $w_{\sigma}=0$ for the curvaton. In
this case the energy density of the curvaton is dominated by its
potential, which in the oscillating epoch roughly takes the form
\begin{eqnarray}
 \rho_\sigma \simeq V(\sigma)
 =2T_3h_{IR}^4+\frac{1}{2}m_{np}^2\sigma^2~,
\end{eqnarray}
where the first term is secondary in the oscillating epoch as will
be explained in the next section.

One can obtain the linear perturbation of the curvaton energy
density in the oscillating epoch as follows,
\begin{eqnarray}
 {\delta\rho_\sigma}_{\rm osc}
 = m_{np}^2\sigma_{\rm osc}\delta\sigma_{\rm osc}~,
\end{eqnarray}
where $\sigma_{\rm osc}$ is the amplitude at the beginning of its
sinusoidal oscillations. It can be viewed as a function of the
curvaton field $\sigma_*$ at Hubble exit, which is the only
relevant quantity since the perturbations from the radiation fluid
is supposed to be negligible. The curvature perturbation is
calculated at the moment of curvaton decay, so it is convenient to
define a parameter
\begin{eqnarray}
 {r_{\sigma}} \equiv
 \frac{3{\rho_{\sigma}}_{\rm dec}}{4{\rho_{\gamma}}_{\rm dec}+3{\rho_{\sigma}}_{\rm dec}}~,
\end{eqnarray}
which characterizes the fraction of curvaton component when it
starts to decay. Consequently, at fixed $\rho_{\rm osc}$ and
$\rho_{\rm dec}$, we obtain the curvature perturbation at linear
order as follows,
\begin{eqnarray}\label{zeta}
 \zeta_{g}\simeq\frac{{r_{\sigma}}H_*}{3\pi\sigma_{\rm osc}}~,
\end{eqnarray}
where we have applied the result presented in Eq. (\ref{delta
sigma}) as well as the approximate relation ${\sigma_{\rm
osc}}_{,\sigma_*}\simeq1$ when the oscillation starts with
$H=m_{np}$.

\subsection{Non-Gaussianities}

In the above section we have studied the primordial perturbations
at linear order. It is necessary to extend the theoretical
framework of this model beyond the leading order. Particularly,
the investigation of non-Gaussianities may provide a powerful
discriminator among numerous models describing the early universe.
For example, non-Gaussianities in a single canonical field
inflation model were calculated in pioneering works
\cite{Acquaviva:2002ud, Maldacena:2002vr} and found to be very
small. Later, studies of DBI inflation
models\cite{Alishahiha:2004eh} predicted that a large
non-Gaussianity of equilateral type can be realized in
inflationary cosmology\cite{Chen:2006nt}. Recent developments in
cosmological perturbation theory shows that a sizable
non-Gaussianity of local type can be obtained in bounce cosmology
but that it takes on negative values\cite{Cai:2009fn, Cai:2009rd}.

\subsubsection{Local type}

To expand the perturbations to second order, we are able to
investigate the features of non-Gaussianities in the curvaton
brane scenario. At first glance, we consider the local form. The
relative magnitude of the nonlinear perturbation is conventionally
specified
by a parameter $f_{\rm NL}$, which is defined by
\begin{eqnarray}
 f_{\rm NL}\equiv\frac{5}{3}\frac{\zeta-\zeta_g}{\zeta_g^2}~.
\end{eqnarray}

In this case, it is convenient to use the $\delta{\cal N}$
formalism to study the non-Gaussian perturbations by virtue of the
separate universe assumption\cite{Sasaki:1995aw, Lyth:2005fi,
Starobinsky:1986fxa}. Starting from an initial flat slice at time
$t_{\rm osc}$, the curvature perturbation can be generally
expanded as
\begin{eqnarray}\label{deltaN}
 \zeta(t,\vec{x})
 = {\cal N}_{,\sigma}\delta\sigma+\frac{1}{2}{\cal N}_{,\sigma\sigma}\delta\sigma^2+...~,
\end{eqnarray}
where ${\cal N}(t,\vec{x})\equiv\ln[{a_{\rm dec}}/{a_{\rm osc}}]$
is the amount of expansion to the final slice of uniform energy
density at curvaton decay. To compare the first term in the r.h.s.
of Eq. (\ref{deltaN}) with the linear curvature perturbation
obtained in Eq. (\ref{zeta}), we can read ${\cal
N}_{,\sigma_*}={2r_{\sigma}}/{3\sigma_{\rm osc}}$. Differentiating
this term with respect to $\sigma_*$ again and applying the
definition of $f_{\rm NL}$, we can obtain
\begin{eqnarray}
 f_{\rm NL} = \frac{5}{6}\frac{{\cal N}_{,\sigma_*\sigma_*}}{{\cal N}_{,\sigma_*}^2}
 \simeq -\frac{5}{3} -\frac{5r_{\sigma}}{6}
 +\frac{5}{4r_{\sigma}}~.
\end{eqnarray}
This form precisely agrees with the result obtained in a normal
curvaton model\cite{Lyth:2002my}, and more detailed studies of this
parameter were present in  \cite{Lyth:2005fi, Huang:2008ze,
Huang:2008bg}.

\subsubsection{Equilateral type}

The above result shows that the local non-Gaussianity in our model
nicely coincides with that given in a general curvaton model. For
this case, we are unable to differentiate the model of curvaton
brane from the standard curvaton model by observing the local
non-Gaussianity. So we need to study more details of the nonlinear
perturbations generated in our model. In particular, we are
interested in the non-Gaussianity of equilateral type.

The presence of interactions in the perturbation Lagrangian leads
to non-Gaussianities. To start, we define the power spectrum
$P_{\zeta}$ and bispectrum $B_{\zeta}$ as follows,
\begin{eqnarray}
\langle {\zeta}_{k_1} {\zeta}_{k_2} \rangle &=& (2\pi)^3
\delta^3(\vec{k}_1+\vec{k}_2) P_{\zeta}(k_1)~,\\
\langle {\zeta}_{k_1} {\zeta}_{k_2} {\zeta}_{k_3} \rangle &=&
(2\pi)^3 \delta^3(\vec{k}_1+\vec{k}_2+\vec{k}_3) \nonumber\\
&&\times B_{\zeta}(k_1,k_2,k_3)~,
\end{eqnarray}
and then these two spectra can be related in terms of a
$k$-dependent nonlinearity parameter $f_{NL}$,
\begin{eqnarray}\label{fnlbasic}
B_{\zeta}(k_1,k_2,k_3) = \frac{3}{10}(2\pi)^4\frac{\sum
k_i^3}{\prod k_i^3}P_{\zeta}^2 f_{NL}(k_1,k_2,k_3)~,
\end{eqnarray}
in momentum space.

It is easiest to work in the interaction picture, in which the
three-point correlator to leading order is given by
\begin{eqnarray}\label{zeta^3}
 \langle {\zeta}_{k_1} {\zeta}_{k_2} {\zeta}_{k_3} \rangle|_{\rm dec}
 &=& i\int_{t_i}^{t_{\rm dec}}dt' \langle [{\zeta}_{k_1} {\zeta}_{k_2} {\zeta}_{k_3}, L_3(t')]
 \rangle \nonumber\\
 &=& i{\cal N}_{,\sigma_*}^3\int_{t_i}^{t_{*}}dt'
 \langle [{\delta\sigma}_{k_1} {\delta\sigma}_{k_2} {\delta\sigma}_{k_3}, L_3(t')]
 \rangle~,\nonumber\\
\end{eqnarray}
where the $\delta {\cal N}$ formalism was applied. For a generic
curvaton model the integral stops at the moment of curvaton decay.
Here $t_i$ corresponds to the initial time before which there are
any non-Gaussianities. The square parentheses indicate the
commutator, and $L_3$ is the interaction Lagrangian which will be
performed in the following. In the second expression of Eq.
(\ref{zeta^3}) we have neglected the integral from the Hubble exit
moment to the curvaton decay, due to the assumption of a curvaton
model that the number of efolds from the end of inflation to the
beginning of the oscillations is completely unperturbed.

If we have computed the three point correlators of curvaton field
fluctuations, the nonlinearity parameter can be obtained by making
use of the above equations. With an assumption of weak coupling
between the fields, we perturb the sound speeds in the quadratic
lagrangian and then obtain the lagrangian with the leading order
terms up to cubic parts,
\begin{eqnarray}
{\cal L}_3 \supseteq \frac{a^3}{2{c_s}^5\dot\sigma}
[\delta\dot\sigma^3-\frac{{c_s}^2}{a^2}\delta\dot\sigma(\nabla\delta\sigma)^2]~.
\end{eqnarray}
Correspondingly, the dominant terms in the interaction Hamiltonian
in Fourier space are given by
\begin{eqnarray}\label{Hint}
H_{int} \supseteq \int dk^3 \bigg[ -\frac{a^3}{2{c_s}^5\dot\sigma}
(\delta\dot\sigma^3+\frac{{c_s}^2}{a^2}k^2\delta\dot\sigma\delta\sigma^2)
\bigg].
\end{eqnarray}
Then we decompose the field fluctuations in canonical quantization
process,
\begin{eqnarray}\label{fieldpert}
 \delta{\sigma}_{k}(t) &=& u(\vec{k}){a}_{\vec k}+u^*(-\vec{k}){a}_{-\vec k}^\dag~,\nonumber\\
 u(\vec{k})&=&\frac{H}{\sqrt{2k^3}}(1+i{c_s}k\tau)e^{-i{c_s}k\tau}~,
\end{eqnarray}
with the creation and annihilation operators defined by $[a_{\vec
k},a_{\vec k'}^\dag]=\delta({\vec k}-{\vec k'})$.

Since we have obtained the interaction Hamiltonian and the modes
of the field fluctuations, we are now able to calculate the three
point correlator of equilateral form,
\begin{eqnarray}
 \langle \delta{\sigma}_{k}^3 \rangle
 &=& -i \int dt \langle [\delta{\sigma}_{k}^3,H_{int}] \rangle \nonumber\\
 &\sim& -\frac{H_*^5}{c_s^2\dot\sigma_*
 k^6}(2\pi)^3\delta(\sum\vec{k}_i)~.
\end{eqnarray}
From Eqs. (\ref{Hint}) and the above result, we can see that
$H_{int}\sim 1/{c_s}^{2}$. This roughly agrees with the result in
usual single DBI inflation which is proportional to
$1/c_s^2$\cite{Chen:2006nt}. To be explicit, we combine Eqs.
(\ref{zeta}), (\ref{fnlbasic}) and (\ref{zeta^3}) and eventually
have the non-Gaussianity parameter of equilateral type,
\begin{eqnarray}
 f_{\rm NL} \simeq -\frac{\sigma_{osc}H_*}{2r_\sigma
 c_s^2h_{IR}^2\sqrt{T_3}}~.
\end{eqnarray}

From the above analysis, we learn that the non-Gaussianity of
equilateral type in the model of curvaton brane can be amplified
both by small $r_\sigma$ and $c_s^2$. We may understand this
phenomenon as follows. The amplification of non-Gaussianities from
$r_\sigma$ is inherited from the physical feature of a general
curvaton model. Moreover, when those perturbation modes exit the
Hubble radius during inflation, the sound speed is also small due
to the relativistic motion of the curvaton brane in the warped
throat, and thus leads to significant equilateral non-Gaussianity
signals.

Note that in the above computations we assumed that the curvaton
enters the oscillating epoch very soon after inflation. So an
approximate relation ${\sigma_{\rm osc}}_{,\sigma_*}\simeq1$ can
be used to obtain an estimate. Under this estimate there is
\begin{eqnarray}\label{sigma osc}
\sigma_{osc}\simeq 5h_{IR}^2\sqrt{T_3}/m_{np}~,
\end{eqnarray}
and so $f_{\rm NL}\simeq-5H_*/2r_{\sigma}c_s^2m_{np}$. It is worth
extending the above analysis to more general cases.

\section{Consistency Relationship and Constraints}

In this section, we shall discuss the consistency conditions and
observational constraints on the phase space of the parameters
appearing in our model.

To our current knowledge, the most powerful observational
constraints on models of early universe are provided by the
CMB experiments. In order to link the observations with the
theoretical studies, let us make a short summary of the observable
predictions obtained in our model. In our model, the amplitudes of
the scalar and tensor perturbation spectra are given by
\begin{eqnarray}
 P_\zeta=\frac{r_\sigma^2m_{np}^2H_*^2}{225\pi^2h_{IR}^4T_3}~,~~
 P_T=\frac{2H_*^2}{\pi^2m_{pl}^2}~,
\end{eqnarray}
respectively, where $m_{pl}=1/\sqrt{8\pi G}$. The corresponding
spectral indices are given by
\begin{eqnarray}
 n_\zeta=1-2\epsilon_*+3s_*~,~~
 n_t=-2\epsilon_*~,
\end{eqnarray}
and the spectral tilt of the scalar perturbation spectrum is the
same as that of the iso-curvature modes during inflation as shown
in Eq. (\ref{n sigma}). The tensor-to-scalar ratio takes the form,
\begin{eqnarray}
 r_T \equiv \frac{P_T}{P_\zeta}
  = \frac{450h_{IR}^4T_3}{r_\sigma^2m_{np}^2m_{pl}^2}~.
\end{eqnarray}
Going beyond the Gaussian statistics in the scalar perturbation
spectrum, the non-Gaussian fluctuations are characterized in terms
of the nonlinearity parameter $f_{\rm NL}$. In the local and
equilateral limits, the leading order contributions to this
parameter are expressed as,
\begin{eqnarray}
 f_{\rm NL}^{\rm local} \simeq -\frac{5}{3}-\frac{5r_\sigma}{6}+\frac{5}{4r_\sigma}~,~~
 f_{\rm NL}^{\rm equil} \simeq -\frac{5H_*}{2r_\sigma{c_s}_*^2m_{np}}~,
\end{eqnarray}
respectively.

Since in our model the position of the curvaton brane is fixed at
the tip of the warped throat, it suffers from very few theoretical
constraints compared to usual brane inflation models, namely, the
bounds from the throat\cite{Baumann:2006cd} and bulk
volume\cite{Chen:2006hs} and the primordial tensor
modes\cite{Lidsey:2007gq} can be circumvented delicately. However,
one still needs to be careful of the local gravitational
backreaction of the probe brane on the throat. In the usual case
this effect is negligible provided
$4\pi{g}_s\alpha'^2\ll{c}_s^2L^4$\cite{Easson:2007dh}, and so
gives a bound on the sound speed of the curvaton brane
${c_s}_*^2\gg0.6/MK$ during inflation.

Combining the Wilkinson Microwave Anisotropy Probe five year
(WMAP5) data \cite{Komatsu:2008hk} with the distance measurements
from the Type Ia supernovae (SN) \cite{Kowalski:2008ez} and the
Baryon Acoustic Oscillations (BAO) in the distribution of galaxies
\cite{Eisenstein:2005su}, the parameters of primordial
cosmological perturbations have been accurately determined by a
group of stringent limits as follows:
$P_\zeta=(2.445\pm0.096)\times10^{-9}$, $n_\zeta=0.96\pm0.013$ at
the pivot scale $k=0.002{\rm Mpc}^{-1}$, $r_T<0.22$ at the
$2\sigma$ level with a prior assumption of no running of the
spectral index, and further $-9<f_{\rm NL}^{\rm local}<111$ and
$-151<f_{\rm NL}^{\rm equil}<253$ at the $2\sigma$ level for the
local and equilateral limits, respectively.

Moreover, the number of efoldings of inflation is also constrained
by cosmological observations. However, the reheating process after
the primordial epoch is quite model-dependent and critically
dependent on the reheating temperature. Up to now, the allowed
reheating temperature can be anywhere between $1{\rm MeV}$ and
$10^{16}{\rm GeV}$ in principle. So it makes the measurement of
the inflationary efolding number undetermined. In the current
work, we take the simplest example of instantaneous curvaton
reheating where the curvaton decays to standard-model particles
momentarily once $H=\Gamma$. Therefore, we simply assume the
inflationary efolding number takes the value ${\cal N}_{\rm
inf}=50$, which is a quite reasonable value allowed by
observations. We would like to leave the more detailed study of
the reheating process of the curvaton brane model to future
studies.


Additionally, there is one extra theoretical constraints from the
requirement that the scalar perturbations generated by the
inflaton field should be negligible during the period of curvaton
oscillations.
We assume that the inflaton field satisfies the traditional
slow-roll conditions, and hence the curvature perturbation during
inflation are given by the power spectrum
$P_S=\frac{H_*^2}{8\pi^2m_{pl}^2\epsilon_*}$.
Similar to the definition of the tensor-to-scalar ratio, one can
define a new parameter
\begin{eqnarray}
 r_S\equiv\frac{P_S}{P_\zeta}
 =\frac{225h_{IR}^4T_3}{8r_\sigma^2m_{np}^2m_{pl}^2\epsilon_*}~,
\end{eqnarray}
which has to satisfy the condition $r_S\ll1$.

\begin{figure}[htbp]
\includegraphics[scale=0.6]{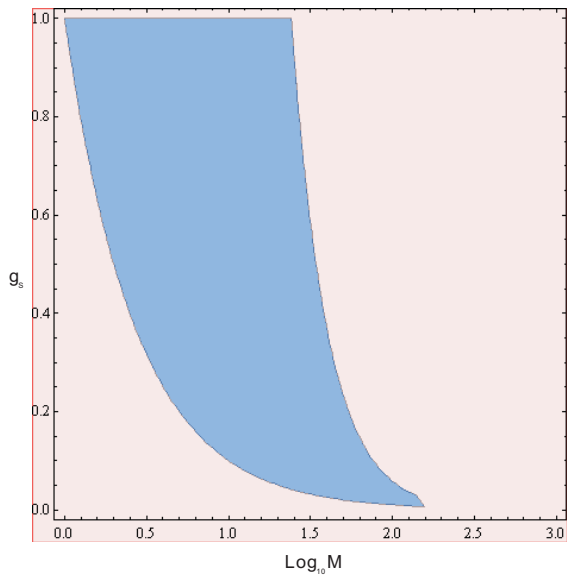}
\caption{ Two dimensional constraints on $g_s$ and $M$ from
current observations, assuming a curvaton-dominated universe after
inflation. The blue region is allowed by current observations. In
the numerical computation, $MK=10^3$ and ${\cal N}=50$ are chosen
as priors. } \label{fig:a}
\end{figure}

\begin{figure}[htbp]
\includegraphics[scale=0.6]{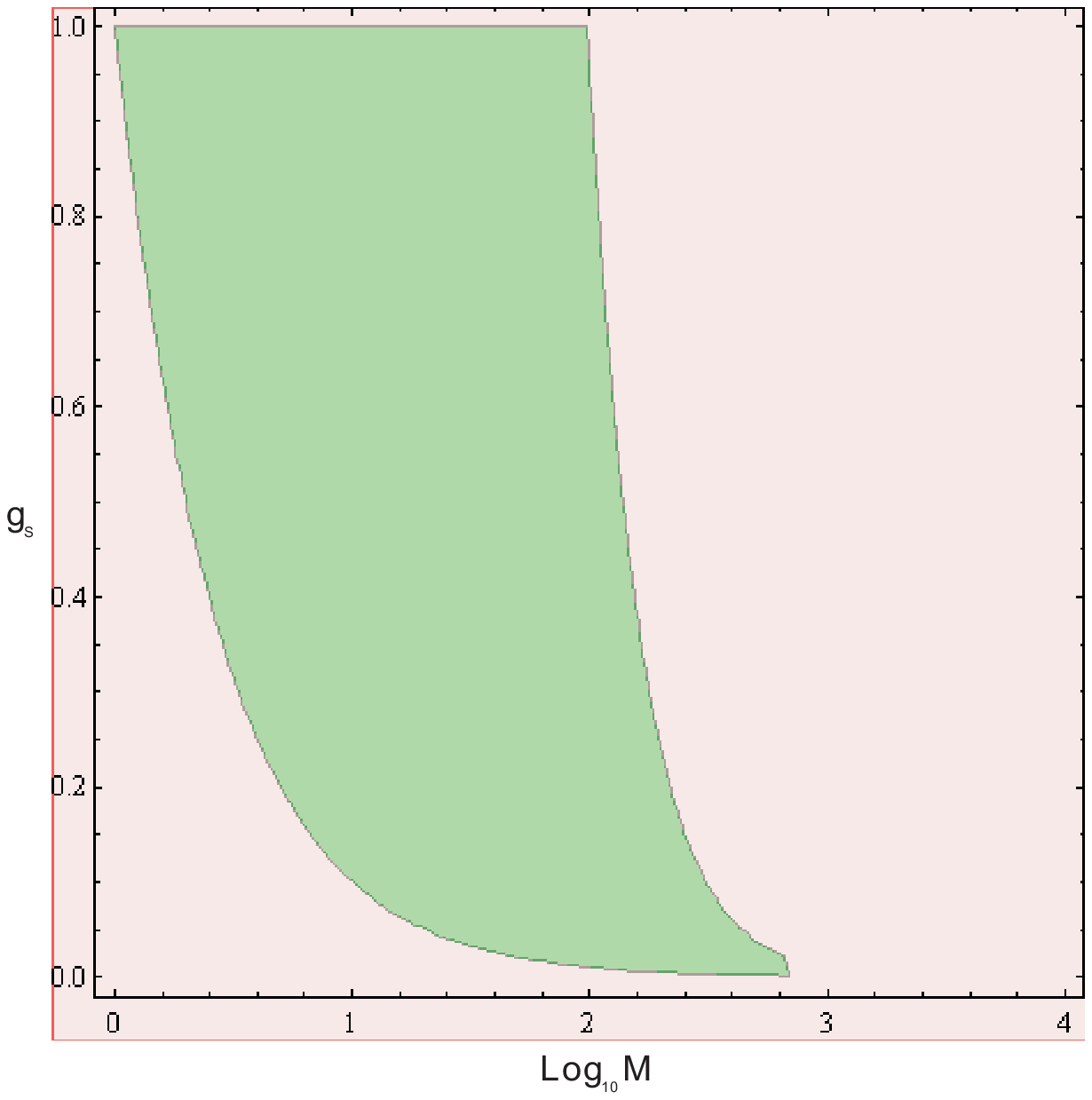}
\caption{ Two dimensional constraints on $g_s$ and $M$ from
current observations, assuming a curvaton-dominated universe after
inflation. The green region is allowed by current observations. In
the numerical computation, $MK=10^4$ and ${\cal N}=50$ are chosen
as priors. } \label{fig:b}
\end{figure}

\begin{figure}[htbp]
\includegraphics[scale=0.6]{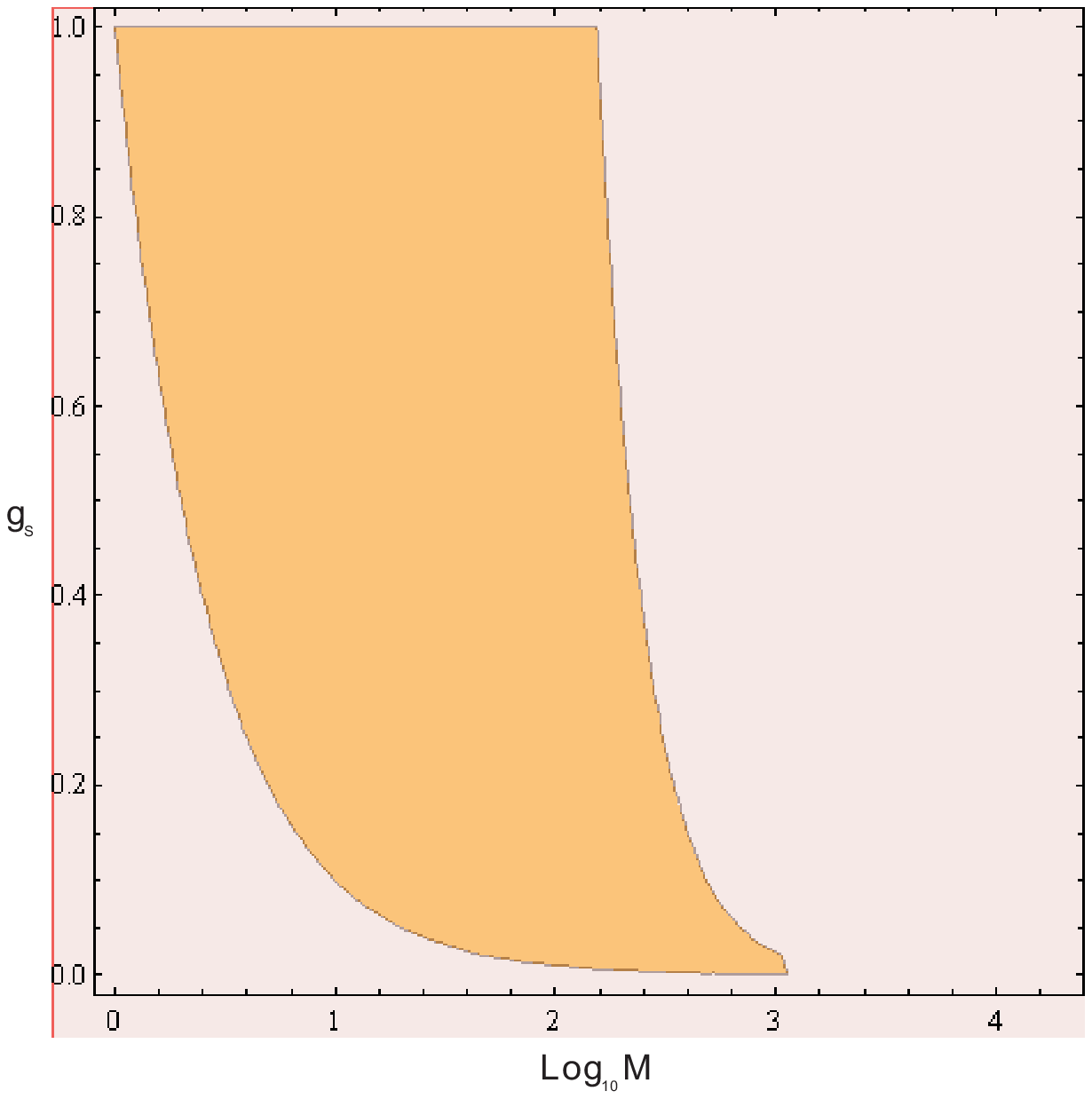}
\caption{ Two dimensional constraints on $g_s$ and $M$ from
current observations, assuming a curvaton-dominated universe after
inflation. The orange region is allowed by current observations.
In the numerical computation, $MK=2\times10^4$ and ${\cal N}=50$
are chosen as priors. } \label{fig:c}
\end{figure}

Taking into account all the conditions introduced above, we find
the available stringy parameters most stringently constrained by
observations are $g_s$ and $M$. In Fig. \ref{fig:a}, \ref{fig:b}
and \ref{fig:c} , we present our numerical calculations of the
permitted regime for $g_s$ and $M$ by fixing the total charge
number of the background $MK$. Our results imply that the allowed
phase space of the $g_s$-$M$ plane with a large $MK$ is much
bigger than that with a small one. This feature is similar to that
of a traditional DBI inflation model, where a large value of $MK$
can provide a small enough warp factor; however, the limits on
stringy parameters are much weaker in our case. This difference is
mainly due to the particular feature of our model that the
curvaton position keeps still during the epoch of inflation.

In addition to the constraints on the stringy parameters performed
above, we can also obtain the
constraint on the slow-roll parameter $\epsilon_*$ as
$0.0174<\epsilon_*<0.0226$. Moreover, in order to make sure that the
equilateral non-Gaussianity generated in our model is consistent
with the WMAP5 data, the fraction of curvaton density $r_\sigma$
in the oscillating epoch needs to be of order $O(1)$. This is
self-consistent with the previous assumption that the universe is
dominated by the curvaton field after inflation.

To conclude, in this section we have confronted our model with the latest
cosmological observations. Under the priors of the inflationary
efolding number and the background charge number, we derived
that there is a
sizable allowed region for the phase space of this model. It
illustrates that cosmological observations can be nicely satisfied
in our model. However, it is crucial to perform a global analysis
which combines both the stringy parameters appearing in the model
and the standard cosmological parameters without any artificial
assumptions. We will leave this work to a future study.

\section{Discussion and Conclusions}

In this paper, we have presented a curvaton model obtained by considering
a $\overline{D3}$-brane in a KS throat with approximate
isometries from string theory. In this model, a
$\overline{D3}$-brane is stretched out along the 4D external
space, and can be localized at the tip of the KS throat because of
the gauge forces. Adding the nonperturbative effect of the
K$\ddot{a}$hler moduli stabilization, its angular coordinates can
play a role of the curvaton while the other degrees of freedom
(for example, the radial motion of the $\overline{D3}$-brane) are
almost frozen. This scenario can be embedded into the framework of
inflation models in terms of multiple branes, in which the
$\overline{D3}$-brane rotates relativistically.

Based on the formulae developed in Ref. \cite{Cai:2008if} and
\cite{Cai:2009hw}, we have studied the background evolution and
the primordial perturbations of this curvaton model. The solution
of the background dynamics of the curvaton is consistent with the
assumptions of the nonperturbative potential of moduli stabilization,
and so ensures the feasibility of our model. The process of
converting the primordial fluctuations of the curvaton into
curvature perturbations takes place after inflation, and is
similar to what happens in the usual curvaton mechanism. We would
like to emphasize that the final curvature perturbations are
calculated under the assumption that the curvaton decays
immediately after the oscillation period ceases. Such a fast decay
rate is also suggested by the stringy reheating configuration, in
which one needs to turn on certain SUSY-breaking deformations of
the KS background\cite{Berndsen:2008my} in order to avoid
redundant long-lived Kaluza-Klein (KK) modes\cite{Kofman:2005yz,
Dufaux:2008br} and have a correct relic density for DM
candidates\cite{Berndsen:2008my, Frey:2009qb}. However, the
precise duration of the decay process is not addressed in the
current work. This would alter the efficiency of the conversion of
curvature perturbations and the precise determination of the
inflationary efolding number.

Our model is related to many other interesting issues which
deserve intensive studies in the future. For example, as mentioned
below Eq. (\ref{S2}), this model is expected to possess nontrivial
kinetic couplings at fourth order from the geometric brane
deformation. Therefore it may leave important signals on the
trispectra which would render the model distinguishable from other
multi-field inflation models with non-standard kinetic
terms\cite{RenauxPetel:2008gi, Gao:2009gd, Mizuno:2009cv,
Byrnes:2009qy, Gao:2009at, Mizuno:2009mv, Battefeld:2009ym,
Chen:2009zp}. In addition, a lesson from string theory suggests
that a more reasonable expression of the multi-brane dynamics ought to
be described by a matrix in the framework of nonabelian
background\cite{Taylor:1999pr}, which shows a potential connection
between the multi-brane scenario and the matrix inflation model
investigated recently\cite{Ashoorioon:2009wa, Berndsen:2009ww}.

\section*{Acknowledgements}
We would like to thank Aaron Berndsen, Robert Brandenberger and
Xingang Chen for helpful discussions and comments on our work.
Y.F.C. acknowledges Prof. Xinmin Zhang for extensive support of his
research. Y.F.C. also thanks the Canadian Institute for Theoretical
Astrophysics, McGill University, Perimeter Institute, Simon Fraser
University, and the University of British Columbia for
hospitality. The research of J.Z. and Y.S.P is supported in part
by NSFC under Grant No:10775180, in part by the Scientific
Research Fund of GUCAS (NO.055101BM03), in part by CAS under Grant
No: KJCX3-SYW-N2.

\end{document}